\documentclass[12pt]{article}
\newcommand{\beq}{\begin{eqnarray}}
\newcommand{\eeq}{\end{eqnarray}}

\newcommand{\be}{\begin{equation}}
\newcommand{\ee}{\end{equation}}
\newcommand{\ben}{\begin{eqnarray}\displaystyle}
\newcommand{\een}{\end{eqnarray}}

\newcommand{\sectiono}[1]{\section{#1}\setcounter{equation}{0}}

\def\sqr#1#2{{\vcenter{\vbox{\hrule height.#2pt
         \hbox{\vrule width.#2pt height#1pt \kern#1pt
            \vrule width.#2pt}
         \hrule height.#2pt}}}}

\begin{document}

{}~ \hfill\vbox{\hbox{hep-ph/0401115} \hbox{PUPT-2108} }\break

\vskip 0.5cm

\begin{center}
\Large{\bf Prospects for Pentaquark Production at Meson Factories}

\vspace{10mm}

\normalsize{Thomas E. Browder,} 

\vspace{2mm}

\normalsize{\em Department of Physics}

\vspace{0.2cm}
\normalsize{\em University of Hawaii, Honolulu, HI 96822}

\vspace{7mm}

\normalsize{Igor R. Klebanov and Daniel R. Marlow }

\vspace{2mm}
\normalsize{\em Joseph Henry Laboratories, Princeton University,}

\vspace{0.2cm}

\normalsize{\em Princeton, NJ 08544}
\end{center}

\vspace{10mm}

\begin{abstract}

\medskip

Following Rosner [hep-ph/0312269], we consider $B$-decay production channels
for the exotic $I=0$ and $I=3/2$ pentaquarks that have been recently
reported. We also discuss new search channels
for isovector pentaquarks, such as the $\Theta^{*++} (\bar s duuu)$, that
are generically present in chiral soliton models but were not
observed in recent experiments.
Futhermore, we argue that weak decays of charmed baryons, such as 
the $\Lambda_c^+$ and $\Xi_c^0$, 
provide another clean way of detecting exotic baryons made
of light quarks only. We also discuss discovery channels for 
charmed pentaquarks,
such as the isosinglet $\Theta_c^0 (\bar c udud)$, in 
weak decays of bottom mesons 
and baryons. Finally, we discuss prospects for inclusive production
of pentaquarks in $e^+ e^-$ collisions, with associated production of 
particles carrying the opposite baryon number. 

\end{abstract}

\newpage

\section{Introduction}

Recently a great deal of experimental and theoretical activity has been 
dedicated to exotic pentaquark baryons.
This activity was sparked by observations of a  narrow
$S=1$ baryon resonance with a mass of
1540 MeV 
\cite{Nakano,Barmin,Stepanyan,Barth,Asratyan,Kubarovsky,Hermes,Aleev}. 
At present the
spin, parity and magnetic moment of this state have not been
determined; some groups, e.g. the SAPHIR collaboration \cite{Barth}, find
that the isospin of the $\Theta^+$ is zero. Such a state
appears naturally in the rigid rotator approach to the 
3-flavor Skyrme model \cite{Manohar,Chemtob,Praszalowicz,DPP}\footnote{
The rigid rotator Skyrmion \cite{Skyrme}
approach to non-exotic baryons is usually
justified in the limit of large $N_c$, where $N_c$ is 
the number of colors \cite{Witten,Adkins}.
However, the rigid rotator
approach to the exotic baryons cannot be justified by large $N_c$
reasoning \cite{KK,IRK,Cohen,US,Cohennew}.
A study of kaon resonances near a Skyrmion, which
is a better approach to the exotics for large $N_c$, shows that
a state with the quantum numbers of $\Theta^+$ does not appear 
for conventional values of model parameters \cite{US}.
However, such a state can be made to appear by large changes
in the parameters.}
or in diquark models \cite{Karliner:2003sy,JW,JWnew}. 
Both of these approaches produce
exotic baryons in the $\bf{\overline {10}}$ of $SU(3)$, with positive parity. 
In the diquark model
the $\bf{\overline {10}}$ 
mixes strongly with another exotic multiplet, ${\bf 8}$ \cite{JW,JWnew};
in the chiral soliton model there is also an ${\bf 8}$ 
of excited baryons coming from the 
breathing modes of the Skyrmion, and there is also some mixing between the 
$\bf{\overline{10}}$ and the $\bf 8$ \cite{Weigel}. 
Therefore, the $SU(3)$ rigid rotator Skyrme model and 
the diquark models give low-lying pentaquark spectra that are 
rather similar, although 
they may be different quantitatively. In particular,
both approaches predict the existence of an $I=3/2$ multiplet of exotics,
which are the lowest strangeness ($S=-2$) members of the antidecuplet.

Quite recently, an experimental boost to these ideas came
from the NA49 collaboration which reported observation of a
manifestly exotic
member of this quartet, $\Xi_{3/2}^{--}(ssdd\bar u) $
\cite{Alt}. For analysis of the theoretical situation in light of this
new result, see \cite{DP,JWnew}.
In spite of the report from NA49, 
more experimental and theoretical work is clearly needed.
The experiments carried out so far are rather difficult,
and care needs to be exercised in interpreting the data
(see, for instance, \cite{krefl} and \cite{Wenig}).
A surprising feature of the experimental results is that the ratio of
the $\Theta^+$ production cross-section to that of $\Lambda (1520)$
is found to be very high, in the range of $30-50\%$ 
(however,
the experimental consistency of a ratio this high was questioned in
\cite{Wenig}). 
Theoretically, the production 
cross-section of a 5-quark state 
is expected to be roughly $O(1\%)$ of that for a 3-quark 
state.\footnote{We thank G. Farrar for pointing this out to us.}

A further question concerns parity of the pentaquarks.
While early quark model approaches 
predict negative parity (see, for example,  
\cite{Jaffe,Strottman,Gignoux:1987cn,Lipkin:1987sk}),
it was later pointed out that wave functions with positive parity
are energetically 
preferred for certain model interactions \cite{Stancu}.
Soliton and diquark models predict positive parity,
while some recent lattice studies instead suggest negative parity
\cite{Csikor,Sasaki}.

The unexpectedly small width of the
pentaquark resonances is also a puzzle. 
Measurements of the width have been largely limited
by the experimental resolution. Careful analysis of some experiments
indicates a width as small as $1$ MeV
\cite{Nussinov,Arndt,Casher,Krein,Cahn}. This poses a challenge
for all presently available theoretical models,
although there are some ideas on why the width is
considerably smaller than a typical ${\cal O}(100)$ MeV width expected of
a strongly interacting state this much above threshold
\cite{DPP,Pra,Roy,Carlson,MK,Bucc}.

Another important question concerns pentaquarks with
$I=1$ and $I=2$. On general grounds, if the isosinglet
$\bar s udud$ pentaquark has been found, then one suspects that
the $\bar s ud uu$ and $\bar s uuuu$ may be discovered as well, with
somewhat higher masses.  In chiral soliton models
one indeed finds that the existence
of the $I=1$ and $I=2$ pentaquarks is tied to the existence of the
$I=0$ one \cite{Walliser,Kobushkin,Cohen,US}. 
In the $SU(3)$ rigid rotator model the $I=1$ and $I=2$ exotics
belong to the $\bf {27}$ and the
$\bf {35}$, respectively. Their origin is more general, however: all that is
needed to construct exotic states of higher isospin
is an $SU(2)$ collective coordinate \cite{US}
whose existence can be argued on large $N_c$ grounds \cite{Cohen}.
In the simplest diquark model, where
all diquarks are assumed to be antisymmetric in color, spin and
flavor, such states are absent \cite{JW,JWnew}. 
If, however,
one also assumes the presence of quark pairs symmetric in flavor
(pairs of quarks with such quantum numbers are present in decuplet baryons), 
then the $I=1$ pentaquark states of $J=3/2$ and $J=1/2$
appear, just as in the chiral soliton models, and restore
agreement with large $N_c$ considerations. 
The $I=1$ states are typically predicted to lie in the
range $1600-1680$ MeV \cite{Walliser,Kobushkin,Cohen,US}
and should be sufficiently narrow to be
 observable. The highest charge member of this multiplet, $\Theta^{*++}$,
should be observable through decay into $p K^+$. There is no
evidence for such a resonance in available $p K^+$ 
scattering data \cite{Arndt,Jennings}; recent photoproduction
experiments also did not find evidence for this state 
\cite{Stepanyan,Barth,Juengst}. In this paper we propose
further searches for $\Theta^{*++}$.

Beyond the pentaquarks made of the light flavors, it is important to search
for heavy flavored pentaquarks, such as the $\Theta_c^0 (\bar c udud)$.
A natural description of such objects in chiral soliton models is provided
by the bound state approach \cite{CK} where they are 
described by bound states or resonances of flavored mesons and solitons.
In this approach the exotic baryons are distinguished from the conventional
ones by the Wess-Zumino term, which attracts $D$ mesons, but repels
$\bar D$ mesons. Heavy quark symmetry was incorporated into the bound state
approach in \cite{Jenkins}.
As the mass of the meson increases, so does
the tendency to bind, and the $\Theta_c^0$ is typically predicted to lie below
the $\bar D N$ threshold \cite{Oh} (see also \cite{Riska,Rho,US}).
The bottom flavored pentaquark, $\Theta_b^+ (\bar b ud ud)$, is even 
more likely to be bound.\footnote{ 
However, the effective lagrangian for heavy mesons near a Skyrmion
is not known precisely; appreciable changes of parameters can make heavy
flavored pentaquarks unbound.}
In the quark model
approaches of \cite{Stancu,JW} the heavy flavored pentaquarks are
also predicted to be bound (in \cite{JW} the
$\Theta_c^0$ is estimated to lie $\sim 100$ MeV below the $\bar D N$ threshold,
in rough agreement with bound state approach estimates \cite{Oh}).

If the $\Theta_c^0$ lies below the threshold, then it has only weak decay modes,
such as $\Theta_c^0\to p K^0 \pi^-$.
If, however, it lies above the threshold, as predicted by other models 
\cite{KL,Cheung}, then it should appear as a narrow $D^- p$ resonance.
One should keep in mind that some earlier charmed pentaquark searches,
such as the Fermilab E791 searches for $\bar c s u u d$ through its weak
decays into $\phi \pi^- p$ or $K^{*0} K^- p$, were not successful 
\cite{Aitala}. 

Clearly, further experimental input on pentaquarks is 
crucial at this point. 
It is important to find pentaquark production mechanisms that give clean and
unambiguous signatures. 
One possibility first suggested by Rosner \cite{Rosner} is
to search for pentaquarks by using baryonic $B$ decays.\footnote{
For other recent pentaquark search proposals see, for example,
\cite{Randrup,Bleicher,Diehl}.}
Although $B$ decays into a 
baryon and antibaryon typically have branching ratios
$< 10^{-4}$, the large numbers of events accumulated at the $B$-factories
offer excellent opportunities for pentaquark detection that are distinct
from photoproduction or kaon scattering that have been used so far.
In this paper we make further suggestions along the lines of
\cite{Rosner}, and also propose other production mechanisms 
involving decays of heavy flavored baryons.
Since the experimental detection of charged particles is typically
easier than that of neutral particles,
we will focus on signatures that do not require
the detection of an anti-neutron (or neutron)
and that lead to final states with charged particles only.

\sectiono{Pentaquarks in B decays}

One of the suggestions in \cite{Rosner} is that $\Theta^+ (1540)$
may be observable through the decay sequences
\begin{equation}
B^0 \rightarrow \Theta^+ \bar p \rightarrow p \bar p K^0
\end{equation}
\begin{equation}
\bar B^0 \rightarrow \bar \Theta^- p \rightarrow p \bar p \bar K^0
\end{equation}
In practice one observes $K_S$ instead of $K^0$. 
The existence of the $\Theta^+$
and its antiparticle may be confirmed either by examining both the $p K_S$
and $\bar{p} K_S$ invariant mass combinations or by using flavor-tagging.
The $\Theta^+$ pentaquark can also appear in a weak decay of the meson
$\bar B_s^0$, which consists of $b\bar s$ quarks. If the $b$ 
quark decays weakly into $d u\bar u$ then we may find
\begin{equation}
\bar B_s^0 \rightarrow \Theta^+ \bar p \rightarrow p \bar p K^0\ .
\end{equation}

Similarly, one can search for the $I=1$, $I_3=1$ pentaquark $\Theta^{*++}$
in the decay
\begin{equation}
B^+ \rightarrow \Theta^{*++} \bar p \rightarrow p \bar p K^+
\end{equation}
by examining the $p K^+$ invariant mass. Another possibility is
\begin{equation}
B^0 \rightarrow \Theta^{*++} \bar p \pi^-\rightarrow p \bar p K^+ \pi^-
\end{equation}
where again we should examine the $p K^+$ invariant mass. On the
other hand, examination of the $\bar p K^+$ invariant mass should
reveal a peak corresponding to the antiparticle of $\Lambda (1520)$,
which provides a useful normalization.
The first of these modes 
seems most promising as far
as statistics are concerned. For example,
Belle finds $96.4^{+11.2}_{-10.5}$ $p \bar p K^+/78$ fb$^{-1}$ 
versus $11.3^{+4.1}_{-3.4}$ $p \bar p K_S/78$
fb$^{-1}$\cite{Belle_ppbarK0}.
Belle also finds $14.5^{+4.6}_{-4.0}$ $p \bar p K^{*+}/78$ fb$^{-1}$ 
where $K^{*+}\to K_S \pi^+$\cite{Belle_ppbarK0}. 
However, searches for peaks in the $pK$ spectra are not reported
in \cite{Belle_ppbarK0}.
The final state $p \bar p K_S \pi^+$ 
is one suggested by Rosner. 
This final state is flavor-specific (allowing one to distinguish
$p K_S$ and $p \bar{K}_S$); however, the signal to
background is marginal.

To produce the $I=0$ charmed pentaquark $\Theta_c^0$ in a $B$-decay, 
there are modes involving the dominant weak process
$\bar b\to \bar c u \bar d$ \cite{Rosner}:
\begin{equation}
B^+ \rightarrow \Theta_c^0 \bar \Delta^+
\ ,\end{equation}
\begin{equation}
B^0 \rightarrow \Theta_c^0 \bar p \pi^+\ .
\end{equation}
If the $\Theta_c$ is below the $\bar D N$ threshold, then it will decay
weakly into $p K^0 \pi^-$. Otherwise, it will decay strongly
into $\bar D^0 n$ or $D^- p$. Since CLEO reported a large signal for
$B^0\to D^{*-} p \bar p \pi^+$ 
($ 32.3\pm 6.0$ events/9.1 fb$^{-1}$) \cite{CLEO},
we should expect a large signal for $B^0\to D^- p \bar p \pi^+$
as well. This mode can be used to search for $D^- p$ resonances.

Just like the $\Theta^+$, the charmed pentaquark may have counterparts
with $I=1$ and $I=2$. The antiparticle of
a member of the $I=1$ multiplet, $\Theta_c^{* +} (\bar c uuud)$,
may be produced through
\begin{equation}
\bar B^0 \rightarrow \bar\Theta_c^{*-} p \rightarrow D^0 \bar p  p 
\end{equation}
where we assumed that it lies above the $D N$ threshold.
This decay mode has been observed 
(Belle finds $\sim 92\pm 11.5$ $D^0 p \bar p/29$ fb$^{-1}$
and $\sim 19\pm 5$ $D^{*0} p \bar p/29$ fb$^{-1}$)\cite{Belle_ppbarD0},
and now the spectrum of
invariant mass of $\bar p D^0$ needs to be analyzed.

\begin{center}
\begin{table}
\label{bdecay_signature}
\caption{Pentaquark signatures in $B$ meson decay}
\begin{tabular}{ccc} 
Pentaquark & Mode & Experimental Signature \\
\hline
$\Theta^+(1540)$ & $B^0\to p \bar{p} K_S$  & $M(p K_S)$ \\
$\Theta^+(1540)$ & $B^+\to p \bar{p} K_S\pi^+$  & $M(p K_S)$ \\
$\Theta^{*++}$ & $B^+\to p \bar{p} K^+$  & $M(p K^+)$ \\
$\Theta^{*++}$ & $B^0\to p \bar{p} K^+ \pi^-$  & $M(p K^+)$ \\
$\Theta_c^0$ & $B^0\to D^- p \bar{p} \pi^+$ & $M(D^- p)$ \\
 $\bar \Theta_c^{*-}$ & $\bar B^0\to D^0 \bar{p} p $ & $M(D^0\bar p)$ 
\end{tabular}
\end{table}
\end{center}

\begin{center}
\begin{table}
\label{charm_signature}
\caption{Pentaquark signatures in charmed baryon decay}
\begin{tabular}{ccc} 
Pentaquark & Mode & Experimental Signature \\
\hline
$\Theta^+(1540)$ & $\Lambda_c^+\to p K_S \bar{K}_S$  & $M(p K_S)$ \\
$\Theta^{*++}$ & $\Lambda_c^+\to p K^+ K^-$  & $M(p K^+)$ \\
$\Xi_{3/2}^{--}(1860)$ & $\Xi_c^0\to \Xi^-\pi^-\pi^+\pi^+$  & $M(\Xi^-\pi^-)$ \\
$\Xi_{3/2}^{--}(1860)$ & $\Xi_c^0\to\Sigma^-K^-\pi^+\pi^+$& $M(\Sigma^- K^-)$\\ 
$\Xi_{3/2}^{+}$ & $\Xi_c^0\to \Xi^-\pi^-\pi^+\pi^+$  & $M(\Xi^-\pi^+ \pi^+)$ \\
$\Xi_{3/2}^{+}$ & $\Xi_c^0\to \Sigma^+ \bar K_S \pi^-$& $M(\Sigma^+ \bar K_S)$\\ 
\end{tabular}
\end{table}
\end{center}

\begin{center}
\begin{table}
\label{other_signature}
\caption{Pentaquark signatures in $B_s^0$ and $\Lambda_b$ decay}
\begin{tabular}{ccc} 
Pentaquark & Mode & Experimental Signature \\
\hline
$\Theta^+ (1540)$ & $\bar B_s^0\to p \bar{p} K_S$  & $M(p K_S)$ \\
$\Theta_{cs}^0$ & $B_s^0\to D_s^- p \bar{p}\pi^+$  & $M(D_s^- p)$ \\
$\Theta_{c}^0$ & $\Lambda_b\to D^- p D^0$  & $M(D^- p)$\\ 
$\Theta_{cs}^0$ & $\Lambda_b\to D_s^- p D^0$  & $M(D_s^- p)$\\ 
$\Theta_{cs}^-$ & $\Lambda_b\to \bar D^0 \Sigma^- D^+$  & $M(\bar D_0 \Sigma^-)$\\ 
\end{tabular}
\end{table}
\end{center}

One may also try to observe pentaquarks containing a $\bar c$ quark and
an $s$-quark. It is natural to suppose that the lightest such states are
the $\Theta_{cs}^0 (\bar c s u du)$ and 
$\Theta_{cs}^- (\bar c s d du)$, which form an $I=1/2$ doublet
\cite{Lipkin:1987sk,Stancu}.
They could be produced, for example, in decays of $B_s^0 (\bar b s)$:
\begin{equation}
B_s^0 \rightarrow \Theta_{cs}^0 \bar p \pi^+
\ ,\end{equation}
\begin{equation}
B_s^0 \rightarrow \Theta_{cs}^- \bar \Delta^+\ .
\end{equation}
In the Skyrme model, such states could emerge as bound states or
resonances of $D_s^-$ near a Skyrmion.
If these states are above the $D_s N$ threshold of 
$\approx 2910$ MeV, then
they could be detected through strong decays, such as 
$\Theta_{cs}^0 \to D_s^- p$,
$\Theta_{cs}^0 \to D^- \Sigma^+$,
 and 
$\Theta_{cs}^- \to \bar D^0 \Sigma^-$. 
Otherwise only their weak decays are allowed (however, earlier
searches for such states through weak decays have not been
successful \cite{Aitala}).

\sectiono{Pentaquarks in baryon decays}

Another promising way of producing pentaquarks is through weak decays of heavy
flavored baryons. Consider, for example, Cabibbo suppressed
decays of the $\Lambda_c^+ (cud)$. If instead of the dominant
mode $c\to s u \bar d$ we consider Cabibbo suppressed modes that proceed
through $c\to s u\bar s$, then we find the following possible
decay modes where only one additional $q\bar q$ pair is produced:
\begin{equation}
\Lambda_c^+ \rightarrow \Theta^+ \bar K^0 \to p K^0 \bar K^0 \ ,
\end{equation}
\begin{equation}
\Lambda_c^+ \rightarrow \Theta^{*++} K^-\to p K^+  K^- \ .
\end{equation}
Belle published a signal for $\Lambda_c \to p K^+ K^-$ with 
$676\pm 89$ events/78 fb$^{-1}$\cite{Belle_lambdac}.
No signal was shown for $\Lambda_c \to p K_S \bar{K}_S$. 
Rough scaling leads one to expect
${\cal O}(90-100)$ $p K_S \bar{K}_S$ events/78 fb$^{-1}$. 
The signal to background should be excellent.

Another good opportunity is provided by decays of $\Xi_c^+ (csu)$
and $\Xi_c^0 (csd)$. Using the dominant mode $c\to s u \bar d$, we
may have
\begin{equation}
\Xi_c^0\to \Xi_{3/2}^{--} \pi^+ \pi^+ \to \Xi^- \pi^- \pi^+ \pi^+\ .
\end{equation}
This is a large mode, and a sample of events
$ \Xi_c^0\to\Xi^- \pi^- \pi^+ \pi^+ $
was already obtained
by Belle for the study of excited $\Xi_c^*$. This,
or another possible mode 
\begin{equation}
\Xi_c^0\to \Xi_{3/2}^{--} \pi^+ \pi^+ \to \Sigma^- K^- \pi^+ \pi^+\ ,
\end{equation}
provide good
opportunities to confirm the discovery \cite{Alt} of the
$\Xi_{3/2}^{--} (1860)$.

It is interesting that the same events could be used to search for
another manifestly exotic member of the $I=3/2$ quartet,
$\Xi_{3/2}^+ (ssuu\bar d)$,
which is a state not yet observed by NA49 \cite{Alt}. 
Indeed, the following decay sequence is available: 
\begin{equation}
\Xi_c^0\to \Xi_{3/2}^{+} \pi^- \to \Xi^- \pi^- \pi^+ \pi^+\ ,
\end{equation}
and $\Xi_{3/2}^+$ could appear as a peak in the distribution of the 
invariant mass of $\Xi^- \pi^+\pi^+$.
Another possible decay sequence is
\begin{equation}
\Xi_c^0\to \Xi_{3/2}^+ \pi^- \to \Sigma^+ \bar K^0 \pi^- \ ,
\end{equation}
but it is not reported as ``seen'' by
the PDG \cite{PDG}.
Perhaps it is also possible to observe 
\begin{equation}
\Xi_c^0\to \Xi_{3/2}^+ \pi^- \to \Xi^0 \pi^+ \pi^- \to \Lambda \pi^0
\pi^+ \pi^-
\ ,
\end{equation}
where the $\Xi^0$ decays weakly into $\Lambda \pi^0$.

Charm pentaquarks, such as the $\Theta_c^0$ may be produced in weak
decays of the $\Lambda_b (bud)$. Using a Cabibbo suppressed mode
$ b\to c d\bar c$, we may find
\begin{equation}
\Lambda_b \to \Theta_c^0 D^0
\end{equation}
or
\begin{equation}
\Lambda_b \to \Theta_c^{*-} D^+
\end{equation}
where $\Theta_c^{*-} (\bar c uddd)$ is the $I=1$, $I_3=-1$
state. Using instead the dominant mode
$ b\to c s\bar c$, we may find, e.g.,
\begin{equation}
\Lambda_b \to \Theta_{cs}^0 D^0
\end{equation}
or
\begin{equation}
\Lambda_b \to \Theta_{cs}^- D^+\ .
\end{equation}

\sectiono{Pentaquarks in quarkonium decays or in $e^+ e^-$
continuum production}

As suggested in \cite{Rosner}, another possible way of observing
the $\Theta^+$ is through the inclusive reaction
\begin{equation}
J/\psi \to \Theta^+ + X \ .
\end{equation}
However, there is little phase space avilable since
$m(J/\psi)\approx 3097$ MeV, while the minimum
mass of $X$ is $m_N + m_K\approx 1435$ MeV.
It may be better to study instead excited charmonium states, such as
the $\psi(2S)$ whose mass is 3686 MeV. Millions of its decays have
been observed at BES. The typical branching fraction for a
$\psi(2S)$ decay to a 
baryon-antibaryon pair is ${\cal O}(10^{-4})$ \cite{PDG},
thus perhaps it is possible to observe the decay
\begin{equation}
\psi(2S) \to \Theta^+ + X  \to p K^0 + X\ .
\end{equation}

Similar events could be expected in the decay of $b\bar b$
mesons, such as the $\Upsilon(1S)$ (millions of its decays
have been recorded by CLEO \cite{Dytman}). It is heavy enough that it
can decay into charmed pentaquarks.
 For example, one could carry out an inclusive search for
events of the type
\begin{equation}
\Upsilon ( 1S) \to \Theta_c^0+ X \to p D^- +X\ .
\end{equation}
Even if the $\Theta_c^0$ is slightly below the $D N$ threshold, 
there could be some indication for it from a $p D^-$ spectrum
rising toward the threshold.
Otherwise, the decay sequence is
\begin{equation}
\Upsilon ( 1S) \to \Theta_c^0+ X \to p K^0 \pi^- +X\ .
\end{equation}

Finally we mention perhaps the most obvious way of producing pentaquarks:
through $e^+ e^- \to q \bar q$ with subsequent production of 
four light quark-antiquark pairs during fragmentation
(this was briefly discussed in \cite{Armstrong}).
While this method is possible off any resonance, a particularly
convenient data set involves the vast number of 
$e^+ e^-\to q \bar{q}$ events accumulated by
the $B$-factories at the $\Upsilon (4 S)$ resonance.
This method seems particularly well-suited for production of pentaquarks 
containing a $\bar c$ quark, such as in the inclusive reactions
\begin{equation}
e^+ e^- \to \Theta_c^0+ X \to p D^- +X\ ,
\end{equation}
if the $\Theta_c^0$ is above the $\bar D N $ threshold, or
\begin{equation}
e^+ e^- \to \Theta_c^0+ X \to p K^0 \pi^- +X\ ,
\end{equation}
if the $\Theta_c^0$ is below the $\bar D N $ threshold. 
 In these cases, due to the hard fragmentation of the charm quark, 
 one can apply a tight momentum requirement, e.g. $p^*_P>2.5$
 GeV, (where $p^*_P$ is the momentum of the pentaquark candidate in
 the center of mass frame), to reduce backgrounds in the search.

\sectiono{Conclusion}

There is strong motivation to search for the
pentaquarks in decays of heavy hadrons or in $e^+ e^-$
collisions, as suggested in \cite{Rosner, Armstrong}
and in this paper. Possible signatures are given in Tables
1, 2, and 3. Besides confirming the existence of the pentaquarks,
these methods promise a precise determination of their
widths and quantum numbers (spin and parity).
Based on experience so far, their widths are expected to be narrow,
much less than the $ 100$ MeV width that one might naively expect
for a strongly decaying resonance. 
Some heavy flavored pentaquarks may even have weak lifetimes.

One uncertainty for all searches is the lack of a good
understanding of the pentaquark production cross-sections.
Cross-sections for known non-exotic resonances, such as
the $\Lambda (1520)$, should be measured first for calibration.
One then expects pentaquark cross-sections to be at least
of order a percent of that.

\section*{Acknowledgments}
We are grateful to G. Farrar, P. Ouyang and E. Witten for useful
discussions.
This material is based upon work
supported by the National Science Foundation Grants No.
PHY-0243680 and PHY-0140311.
We acknowledge support from the U.S.\ Department of Energy under
grants DE-FG03-94ER40833, DE-FG02-96ER41005.
Any opinions, findings, and conclusions or recommendations expressed in
this material are those of the authors and do not necessarily reflect
the views of the National Science Foundation.

\begingroup\raggedright\endgroup
\end{document}